# Chiral Polyhedra Derived From Coxeter Diagrams and Quaternions

Mehmet Koca[a)], Nazife Ozdes Koca[b)]
and
Muna Al-Shu'eili[c)]

Department of Physics, College of Science, Sultan Qaboos University
P.O. Box 36, Al-Khoud, 123 Muscat, Sultanate of Oman

**Abstract**

There are two chiral Archimedean polyhedra, the snub cube and snub dodecahedron together with their dual Catalan solids, pentagonal icositetrahedron and pentagonal hexacontahedron. In this paper we construct the chiral polyhedra and their dual solids in a systematic way. We use the proper rotational subgroups of the Coxeter groups $W(A_1 \oplus A_1 \oplus A_1)$, $W(A_3)$, $W(B_3)$, and $W(H_3)$ to derive the orbits representing the solids of interest. They lead to the polyhedra tetrahedron, icosahedron, snub cube, and snub dodecahedron respectively. We prove that the tetrahedron and icosahedron can be transformed to their mirror images by the proper rotational octahedral group $W(B_3)/C_2$ so they are not classified in the class of chiral polyhedra. It is noted that vertices of the snub cube and snub dodecahedron can be derived from the vectors, which are linear combinations of the simple roots, by the actions of the proper rotation groups $W(B_3)/C_2$ and $W(H_3)/C_2$ respectively. Their duals are constructed as the unions of three orbits of the groups of concern. We also construct the polyhedra, quasiregular in general, by combining chiral polyhedra with their mirror images. As a by product we obtain the pyritohedral group as the subgroup the Coxeter group $W(H_3)$ and discuss the constructions of pyritohedrons. We employ a method which describes the Coxeter groups and their orbits in terms of quaternions.

[a)] electronic-mail: kocam@squ.edu.om
[b)] electronic-mail: nazife@squ.edu.om
[c)] electronic-mail: m054946@squ.edu.om

# 1 Introduction

It seems that the Coxeter groups and their orbits [1] derived from the Coxeter diagrams describe the molecular structures [2], viral symmetries [3], crystallographic and quasi crystallographic materials [4]. Chirality is a very interesting topic in molecular chemistry and physics. A number of molecules display one type of chirality; they are either left-oriented or right-oriented molecules. In fundamental physics chirality plays very important role. For example a massless Dirac particle has to be either in the left handed state or in the right handed state. No Lorentz transformation exist transforming one state to the other state. The weak interactions which is described by the standard model of high energy physics is invariant under one type of chiral transformations. In three dimensional Euclidean space, which will be the topic of this paper, the chirality is defined as follows: the object which can not be transformed to its mirror image by other than the proper rotations and translations are called chiral objects. For this reason the chiral objects lack the plane and/or central inversion symmetry. In two previous papers we have constructed the vertices of the Platonic–Archimedean solids [5] and the dual solids of the Archimedean solids, the Catalan solids [6], using the quaternionic representations of the rank-3 Coxeter groups. Two of the 13 Archimedean solids, the snub cube and snub dodecahedron are the chiral polyhedra whose symmetries are the proper rotational subgroups of the octahedral group and the icosahedral group respectively.

In this paper we use a similar technique of references [5-6] to construct the vertices of the chiral Archimedean solids, snub cube, snub dodecahedron and their duals. They have been constructed by employing several techniques [7-8] but it seems that the method in what follows has not been studied earlier in this context. We follow a systematic method for the construction of the chiral polyhedra. First we begin with the Coxeter diagrams $A_1 \oplus A_1 \oplus A_1$ and $A_3$ which lead to the tetrahedron and icosahedron respectively and prove that they possess larger proper rotational symmetries which transform them to their mirror images so that they are not chiral solids. We organize the paper as follows. In Sec.2 we construct the Coxeter groups $W(A_1 \oplus A_1 \oplus A_1)$, $W(A_3)$, $W(B_3)$, and $W(H_3)$ in terms of quaternions. In Sec.3 we obtain the proper rotation subgroup of the Coxeter group $W(A_1 \oplus A_1 \oplus A_1)$, and determine the vertices of the tetrahedron by imposing some conditions on the general vector expressed in terms of simple roots of the diagram $A_1 \oplus A_1 \oplus A_1$ . We prove that the tetrahedron can be transformed to its mirror image by the proper octahedral rotation group $W(B_3)/C_2$ . In Sec.4 we discuss similar problem for the Coxeter-Dynkin diagram $A_3$ leading to an icosahedron and again prove that it can be transformed by the group $W(B_3)/C_2$ to its mirror image which indicates that neither tetrahedron nor icosahedron are chiral solids. Here we also discuss the properties of the pyritohedral group and the constructions of the pyritohedrons. The Sec.5 deals with the construction of the snub Cube and its dual pentagonal icositetrahedron from the proper rotational octahedral symmetry $W(B_3)/C_2$ using the same technique employed in Sec.3 and Sec.4. In Sec.6 we repeat a similar work for the constructions of the snub dodecahedron and its dual pentagonal hexacontahedron from the proper icosahedral group $W(H_3)/C_2 \approx A_5$ which is isomorphic to the group of even permutations of five letters. In the concluding Sec.7 we point out that our technique can be extended to determine the chiral polyhedra in higher dimensions.

## 2 Construction of the groups $W(A_1 \oplus A_1 \oplus A_1)$, $W(A_3)$, $W(B_3)$, and $W(H_3)$ in terms of quaternions.

Let $q = q_0 + q_i e_i$, $(i = 1,2,3)$ be a real unit quaternion with its conjugate defined by $\bar{q} = q_0 - q_i e_i$ and the norm $q\bar{q} = \bar{q}q = 1$. The quaternionic imaginary units satisfy the relations

$$e_i e_j = -\delta_{ij} + \varepsilon_{ijk} e_k, \quad (i, j, k = 1,2,3) \tag{1}$$

where $\delta_{ij}$ and $\varepsilon_{ijk}$ are the Kronecker and Levi-Civita symbols and summation over the repeated indices is implicit. The unit quaternions form a group isomorphic to the unitary group $SU(2)$. With the definition of the scalar product

$$(p,q) = \frac{1}{2}(\bar{p}q + \bar{q}p) = \frac{1}{2}(p\bar{q} + q\bar{p}), \tag{2}$$

quaternions generate the four-dimensional Euclidean space. The Coxeter diagram $A_1 \oplus A_1 \oplus A_1$ can be represented by its quaternionic roots in Fig.1 with the normalization $\sqrt{2}$.

$$\bullet \qquad \bullet \qquad \bullet$$
$$\sqrt{2}e_1 \qquad \sqrt{2}e_2 \qquad \sqrt{2}e_3$$

**Figure 1.** The Coxeter diagram $A_1 \oplus A_1 \oplus A_1$ with quaternionic simple roots.

The Cartan matrix and its inverse are given as follows

$$C = \begin{bmatrix} 2 & 0 & 0 \\ 0 & 2 & 0 \\ 0 & 0 & 2 \end{bmatrix}, \quad C^{-1} = \frac{1}{2}\begin{bmatrix} 1 & 0 & 0 \\ 0 & 1 & 0 \\ 0 & 0 & 1 \end{bmatrix}. \tag{3}$$

For any Coxeter diagram, the simple roots $\alpha_i$ and their dual vectors $\omega_i$ satisfy the scalar product [9]

$$(\alpha_i, \alpha_j) = C_{ij}, \quad (\omega_i, \omega_j) = (C^{-1})_{ij}, \quad (\alpha_i, \omega_j) = \delta_{ij}; \quad i, j = 1,2,3. \tag{4}$$

We note also that they can be expressed in terms of each other:

$$\alpha_i = C_{ij}\omega_j, \quad \omega_i = (C^{-1})_{ij}\alpha_j. \tag{5}$$

Let $\alpha$ be an arbitrary quaternionic simple root. Then the reflection of an arbitrary vector $\Lambda$ with respect to the plane orthogonal to the simple root $\alpha$ is given by [10]

$$r\Lambda = -\frac{\alpha}{\sqrt{2}}\bar{\Lambda}\frac{\alpha}{\sqrt{2}} \equiv [\frac{\alpha}{\sqrt{2}}, -\frac{\alpha}{\sqrt{2}}]^*\Lambda. \tag{6}$$

Our notations for the rotary reflections and the proper rotations will be $[p,q]^*$ and $[p,q]$ respectively where $p$ and $q$ are arbitrary quaternions.
The Coxeter group $W(A_1 \oplus A_1 \oplus A_1)$ is generated by three commutative group elements

$$r_1 = [e_1, -e_1]^*, \quad r_2 = [e_2, -e_2]^*, \quad r_3 = [e_3, -e_3]^* . \tag{7}$$

They generate an elementary Abelian group $W(A_1 \oplus A_1 \oplus A_1) \approx C_2 \times C_2 \times C_2$ of order 8. Its proper rotation subgroup elements are given by

$$I = [1,1], \quad r_1 r_2 = [e_3, \bar{e}_3], \quad r_2 r_3 = [e_1, \bar{e}_1], \quad r_3 r_1 = [e_2, \bar{e}_2] . \tag{8}$$

The next Coxeter group which will be used is the tetrahedral group $W(A_3) \approx T_d \approx S_4$. Its diagram $A_3$ with its quaternionic roots is shown in Fig.2.

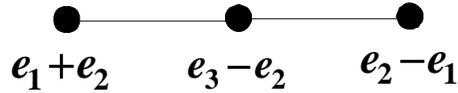

**Figure 2.** The Coxeter diagram $A_3$ with quaternionic simple roots.

The Cartan matrix of the Coxeter diagram $A_3$ and its inverse matrix are given respectively by the matrices

$$C = \begin{bmatrix} 2 & -1 & 0 \\ -1 & 2 & -1 \\ 0 & -1 & 2 \end{bmatrix}, \quad C^{-1} = \frac{1}{4}\begin{bmatrix} 3 & 2 & 1 \\ 2 & 4 & 2 \\ 1 & 2 & 3 \end{bmatrix}. \tag{9}$$

The generators of the Coxeter group $W(A_3)$ are given by

$$r_1 = [\frac{1}{\sqrt{2}}(e_1 + e_2), -\frac{1}{\sqrt{2}}(e_1 + e_2)]^*,$$
$$r_2 = [\frac{1}{\sqrt{2}}(e_3 - e_2), -\frac{1}{\sqrt{2}}(e_3 - e_2)]^*,$$
$$r_3 = [\frac{1}{\sqrt{2}}(e_2 - e_1), -\frac{1}{\sqrt{2}}(e_2 - e_1)]^* \tag{10}$$

The group elements of the Coxeter group which is isomorphic to the tetrahedral group of order 24 can be written compactly by the set [11]

$$W(A_3) = \{[p,\bar{p}] \oplus [t,\bar{t}]^*\}, \quad p \in T, \; t \in T'. \tag{11}$$

Here $T$ and $T'$ represent respectively the binary tetrahedral group of order 24 and the coset representative $T' = O/T$ where $O$ is the binary octahedral group of quaternions of order 48 [11]. The Coxeter diagram $B_3$ leading to the octahedral group $W(B_3) \approx O_h$ is shown in Fig. 3.

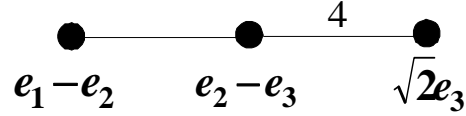

**Figure 3.** The Coxeter diagram $B_3$ with quaternionic simple roots.

The Cartan matrix of the Coxeter diagram $B_3$ and its inverse matrix are given by

$$C = \begin{bmatrix} 2 & -1 & 0 \\ -1 & 2 & -\sqrt{2} \\ 0 & -\sqrt{2} & 2 \end{bmatrix}, \quad C^{-1} = \begin{bmatrix} 1 & 1 & \frac{1}{\sqrt{2}} \\ 1 & 2 & \sqrt{2} \\ \frac{1}{\sqrt{2}} & \sqrt{2} & \frac{3}{2} \end{bmatrix}. \tag{12}$$

The generators,

$$r_1 = [\frac{1}{\sqrt{2}}(e_1 - e_2), -\frac{1}{\sqrt{2}}(e_1 - e_2)]^*, \; r_2 = [\frac{1}{\sqrt{2}}(e_2 - e_3), -\frac{1}{\sqrt{2}}(e_2 - e_3)]^*, \; r_3 = [e_3, -e_3]^* \tag{13}$$

generate the octahedral group which can be written as

$$W(B_3) \approx Aut(A_3) \approx S_4 \times C_2 = \{[p,\bar{p}] \oplus [p,\bar{p}]^* \oplus [t,\bar{t}] \oplus [t,\bar{t}]^*\}, \quad p \in T, t \in T'. \tag{14}$$

A shorthand notation could be $W(B_3) = \{[T,\bar{T}] \oplus [T,\bar{T}]^* \oplus [T',\bar{T}'] \oplus [T',\bar{T}']^*\}$. Note that we have three maximal subgroups of the octahedral group $W(B_3)$, namely, the tetrahedral group $W(A_3)$, the chiral octahedral group consisting of the elements $W(B_3)/C_2 = \{[T,\bar{T}] \oplus [T',\bar{T}']\}$, and the pyritohedral group consisting of the elements $T_h \approx A_4 \times C_2 = \{[T,\bar{T}] \oplus [T,\bar{T}]^*\}$. The pyritohedral symmetry represents the symmetry of the pyritohedrons, an irregular dodecahedron, with irregular pentagonal faces which occurs in iron pyrites.
The Coxeter diagram $H_3$ leading to the icosahedral group is shown in Fig. 4.

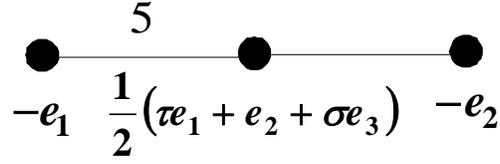

**Figure 4.** The Coxeter diagram of $H_3$ with quaternionic simple roots. (It is assumed that the simple roots are multiplied by $\sqrt{2}$ )

The Cartan matrix of the diagram $H_3$ and its inverse are given as follows:

$$C = \begin{bmatrix} 2 & -\tau & 0 \\ -\tau & 2 & -1 \\ 0 & -1 & 2 \end{bmatrix}, \quad C^{-1} = \frac{1}{2}\begin{bmatrix} 3\tau^2 & 2\tau^3 & \tau^3 \\ 2\tau^3 & 4\tau^2 & 2\tau^2 \\ \tau^3 & 2\tau^2 & \tau+2 \end{bmatrix}. \tag{15}$$

The generators,

$$r_1 = [e_1, -e_1)]^*, \, r_2 = [\tfrac{1}{2}(\tau e_1 + e_2 + \sigma e_3), -\tfrac{1}{2}(\tau e_1 + e_2 + \sigma e_3)]^*, \, r_3 = [e_2, -e_2]^* \tag{16}$$

generate the icosahedral group

$$I_h \approx W(H_3) = \{[p,\bar{p}] \oplus [p,\bar{p}]^*\} \approx A_5 \times C_2, \quad (p, \bar{p} \in I), \tag{17}$$

or shortly, $W(H_3) = \{[I,\bar{I}] \oplus [I,\bar{I}]^*\}$. Here $\tau = \dfrac{1+\sqrt{5}}{2}$, $\sigma = \dfrac{1-\sqrt{5}}{2}$ and $I$ is the set of 120 quaternionic elements of the binary icosahedral group [10]. The chiral icosahedral group is represented by the set $W(H_3)/C_2 \approx A_5 = \{[I,\bar{I}]\}$ which is isomorphic to the even permutations of five letters. Note also that the pyritohedral group is a maximal subgroup of the Coxeter group $W(H_3)$. All finite subgroups of the groups $O(3)$ and $O(4)$ in terms of quaternions can be found in reference [12].

A general vector in the dual space is represented by the vector $\Lambda = a_1\omega_1 + a_2\omega_2 + a_3\omega_3 \equiv (a_1 a_2 a_3)$. We will use the notation $O(\Lambda) = W(G)\Lambda = O(a_1 a_2 a_3)$ for the orbit of the Coxeter group $W(G)$ generated from the vector $\Lambda$ where the letter $G$ represents the Coxeter diagram. We follow the Dynkin notation to represent an arbitrary vector $\Lambda = (a_1 a_2 a_3)$ in the dual space and drop the basis vectors $\omega_i$, $i=1,2,3$. In the Lie algebraic representation theory the components $(a_1 a_2 a_3)$ of the vector $\Lambda$ are called the Dynkin indices [13] which are non-negative integers if it represents the highest weight vector. Here we are not restricted to the integer values of the Dynkin indices. They can be any real number. When the components of the vector in the dual space are non integers values we will separate them by commas otherwise no commas will be used. For an arbitrary Coxeter diagram of rank 3 we define the fundamental orbits as

$$O(\omega_1) = O(100), \ O(\omega_2) = O(010), \ \text{and} \ O(\omega_3) = O(001). \tag{18}$$

Any linear combination of the basis vectors $\omega_i$ over the real numbers will, in general, lead to quasi regular polyhedra under the action of the Coxeter group. In the next four sections we discuss a systematic construction of chiral polyhedra and their dual solids. In our construction tetrahedron and icosahedron will also occur but we prove that they are not chiral polyhedra.

## 3 The orbit $O(\Lambda) = (C_2 \times C_2)(a_1 a_2 a_3)$ as tetrahedron

The proper rotation subgroup of the Coxeter group $W(A_1 \oplus A_1 \oplus A_1)$ applies on an arbitrary vector $\Lambda = \frac{1}{\sqrt{2}}(a_1 e_1 + a_2 e_2 + a_3 e_3)$ as follows:

$$r_1 r_2 \Lambda = \frac{1}{\sqrt{2}}(-a_1 e_1 - a_2 e_2 + a_3 e_3),$$

$$r_2 r_3 \Lambda = \frac{1}{\sqrt{2}}(a_1 e_1 - a_2 e_2 - a_3 e_3),$$

$$r_3 r_1 \Lambda = \frac{1}{\sqrt{2}}(-a_1 e_1 + a_2 e_2 - a_3 e_3).$$

To obtain a tetrahedron from these four vertices the Dynkin indices should satisfy the relations

$$a_1^2 = a_2^2 = a_3^2 = a^2, \quad a_i = \pm a, \ i = 1, 2, 3. \tag{19}$$

We take $a = \frac{1}{\sqrt{2}}$ and start with a vector $\Lambda_I = \frac{1}{2}(e_1 + e_2 + e_3)$ then the orbit $O(\Lambda_I)$ will be given by

$$O(\Lambda_I) = \{\frac{1}{2}(e_1 + e_2 + e_3), \frac{1}{2}(e_1 - e_2 - e_3), \frac{1}{2}(-e_1 - e_2 + e_3), \frac{1}{2}(-e_1 + e_2 - e_3)\}. \tag{20}$$

The tetrahedron with these vertices is shown in Fig.5.

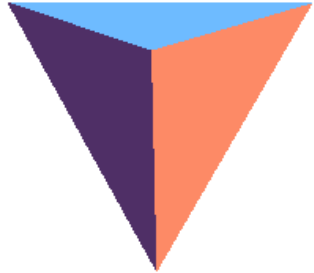

**Figure 5.** The tetrahedron with the vertices given in equation (20).

These are the vertices of a tetrahedron invariant under the rotation group given in (8). Of course the full symmetry of the tetrahedron is a group of order 24 isomorphic to the permutation group $S_4$ generated by reflections of the Coxeter-Dynkin diagram $A_3$ [6]. Now the mirror image of the tetrahedron of (20) can be determined applying the same group of elements in (8) on the vector $r_1 \Lambda_I \equiv \Lambda_{II} = \frac{1}{2}(-e_1 + e_2 + e_3)$. Then the second orbit which is the mirror image of the set in (20) is determined to be

$$O(\Lambda_{II}) = \{\frac{1}{2}(-e_1 - e_2 - e_3), \frac{1}{2}(-e_1 + e_2 + e_3), \frac{1}{2}(e_1 - e_2 + e_3), \frac{1}{2}(e_1 + e_2 - e_3)\}. \quad (21)$$

Of course we know that the union of two orbits in (20) and (21) determines the vertices of a cube. The point here is that if we were restricted to the group $C_2 \times C_2$ of (8) then the tetrahedron in (20) would be a chiral solid. However this is not true because there exist additional rotational symmetries which exchange these two orbits of (20) and (21) proving that the tetrahedron is not a chiral solid. Now we discuss these additional symmetries. It is obvious that the Coxeter diagram in Fig.1 has an additional $S_3$ symmetry which permutes three $A_1$ diagrams. Indeed this symmetry extends the group $C_2 \times C_2$ to the proper octahedral rotation group as will be explained now. One of the generators of this symmetry $d = [\frac{e_2 + e_3}{\sqrt{2}}, -\frac{e_2 + e_3}{\sqrt{2}}]$ is a 2-fold rotation leading to the transformation $e_1 \to -e_1, e_2 \to e_3, e_3 \to e_2$. It is straightforward to see that $dO(\Lambda_I) = O(\Lambda_{II})$. This proves that by a proper rotation tetrahedron can be transformed to its mirror image therefore it is not a chiral solid. The generator $d$ and those elements in (8) enlarge the symmetry to a group of order 8 which can be concisely written as the set of elements $\{[V_0, \bar{V}_0] \oplus [V_1, \bar{V}_1]\}$ [11] where the sets $V_0$ and $V_1$ are defined by

$$V_0 = \{\pm 1, \pm e_1, \pm e_2, \pm e_3\}, \quad V_1 = \{\frac{\pm 1 \pm e_1}{\sqrt{2}}, \frac{\pm e_2 \pm e_3}{\sqrt{2}}\}. \quad (22)$$

A cyclic subgroup $C_3$ of the symmetric group $S_3$ permutes three sets like those in (22) extending the group $C_2 \times C_2$ of order 4 to a group of order 24. Actually the larger group obtained by this extension is the chiral octahedral group of order 24 which can be symbolically written as

$$\{[T, \bar{T}] \oplus [T', \bar{T}']\} \approx S_4 \approx W(B_3)/C_2. \quad (23)$$

This is the proper rotational symmetry of the octahedron whose vertices are represented by the set of quaternions $(\pm e_1, \pm e_2, \pm e_3)$ and the cube whose vertices are the union of the orbits $O(\Lambda_I) \oplus O(\Lambda_{II})$.

**4 The icosahedron derived from the orbit $O(\Lambda) = (W(A_3)/C_2)(a_1 a_2 a_3)$**

The proper rotational subgroup of the Coxeter group $W(A_3) \approx S_4$ is the tetrahedral group $A_4$, the even permutations of the four letters, of order 12. They can be generated by the generators $a = r_1 r_2$ and $b = r_2 r_3$ which satisfy the generation relation $a^3 = b^3 = (ab)^2 = 1$. Let $\Lambda = (a_1 a_2 a_3)$ be a general vector. The following sets of vertices form equilateral triangles

$$(\Lambda, \ r_1 r_2 \Lambda, \ (r_1 r_2)^2 \Lambda), \qquad (\Lambda, \ r_2 r_3 \Lambda, \ (r_2 r_3)^2 \Lambda) \tag{24}$$

with the respective square of edge lengths $2(a_1^2 + a_1 a_2 + a_2^2)$ and $2(a_2^2 + a_2 a_3 + a_3^2)$. We have another vertex $r_1 r_3 \Lambda = r_3 r_1 \Lambda$ as shown in Fig.6.

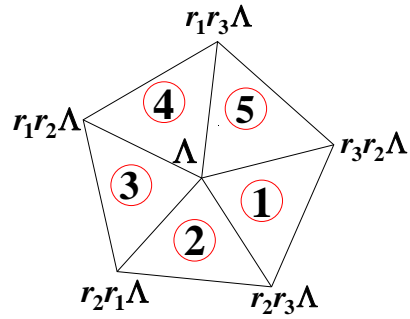

**Figure 6.** The vertices connected to the general vertex $\Lambda$.

Then one can obtain three more triangles by joining $r_1 r_3 \Lambda$ to the vertices $r_1 r_2 \Lambda$ and $r_3 r_2 \Lambda$ and by drawing a line between $r_2 r_1 \Lambda$ and $r_2 r_3 \Lambda$. If we require that all these five triangles be equal each other then we obtain the relations

$$(a_1^2 + a_1 a_2 + a_2^2) = (a_2^2 + a_2 a_3 + a_3^2) = (a_1^2 + a_3^2) . \tag{25}$$

Factoring by $a_2^2$ and defining $x = \dfrac{a_1}{a_2}$ and $y = \dfrac{a_3}{a_2}$ one obtains $x = y$ and a cubic equation $x^3 - x^2 - x = 0$. Assuming $x \neq 0$ we get the solutions $x = \tau$ and $x = \sigma$. This leads to two vectors $\Lambda_I = a_2 (\tau \omega_1 + \omega_2 + \tau \omega_3)$ and $\Lambda_{II} = a_2 (\sigma \omega_1 + \omega_2 + \sigma \omega_3)$. Here $a_2$ is an overall scale factor which can be adjusted accordingly. The $x = 0$ solution represents an octahedron which is not a chiral solid anyway. Let us study the orbit which is obtained from the vector $\Lambda_I$. When expressed in terms of quaternions it will read $\Lambda_I = a_2 \tau (e_2 + \tau e_3)$. We choose the scale factor $a_2 = \dfrac{-\sigma}{2}$ for convenience. To obtain the orbit $O(\Lambda_I) = (W(A_3)/C_2) \Lambda_I$ we use the generators of the tetrahedral group of interest in terms of quaternions, namely,

$$c \equiv [\frac{1}{2}(1+e_1+e_2+e_3), \frac{1}{2}(1-e_1-e_2-e_3)], \quad d \equiv [e_1, -e_1]. \tag{26}$$

They act on the quaternionic units as follows:

$$c: e_1 \to e_2 \to e_3 \to e_1; \quad d: e_1 \to e_1, \ e_2 \to -e_2, \ e_3 \to -e_3. \tag{27}$$

Applying the generators $c$ and $d$ several times on the vector $\Lambda_I = \frac{1}{2}(e_2 + \tau e_3)$ we obtain the set of vectors

$$O(\Lambda_I) = \{\frac{1}{2}(\pm e_1 \pm \tau e_2), \ \frac{1}{2}(\pm e_2 \pm \tau e_3), \ \frac{1}{2}(\pm e_3 \pm \tau e_1)\} \tag{28}$$

which constitute the vertices of the icosahedron shown in Fig. 7.

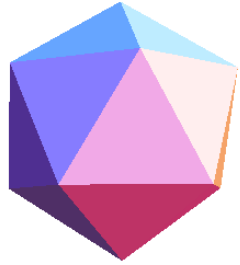

**Figure 7.** The icosahedron obtained from tetrahedral symmetry $A_4$.

Similarly if we use the solution $x = \sigma$ we will get the vector $\Lambda_{II} = \frac{1}{2}(-\tau e_2 + e_3)$ after a suitable choice of the factor $a_2$. Acting the generators in (27) repeatedly on the vector $\Lambda_{II}$ we get the orbit

$$O(\Lambda_{II}) = \{\frac{1}{2}(\pm \tau e_1 \pm e_2), \ \frac{1}{2}(\pm \tau e_2 \pm e_3), \ \frac{1}{2}(\pm \tau e_3 \pm e_1)\}. \tag{29}$$

This is another icosahedron which is the mirror image of the icosahedron of (28). Indeed one can show that $r_1 O(\Lambda_I) = O(r_1 \Lambda_I) = O(\Lambda_{II})$. Before we proceed further we note the fact that the Coxeter-Dynkin diagram $A_3$ has the diagram symmetry $\alpha_1 \leftrightarrow \alpha_3$ and $\alpha_2 \to \alpha_2$, in other words, $\omega_1 \leftrightarrow \omega_3$ and $\omega_2 \to \omega_2$. It is clear that this symmetry does not alter the orbits since under the diagram symmetry $\Lambda_I$ and $\Lambda_{II}$ remain intact. The diagram symmetry acts on the quaternions as $e_1 \to -e_1, \ e_2 \to e_2$ and $e_3 \to e_3$. This transformation can be obtained by the action of the element $[e_1, -e_1]^*$ which is not an element of the group $W(A_3) \approx S_4$. The proper rotation group $A_4 \approx [T, \bar{T}]$ can then be extended by the generator $[e_1, -e_1]^*$ to the pyritohedral group $T_h \approx A_4 \times C_2 = \{[T,\bar{T}] \oplus [T,\bar{T}]^*\}$. Actually in the paper [6] we have shown that the orbit $O(\Lambda_I)$ is invariant under a larger group $I_h$ of (17) which admits the pyritohedral group as a maximal subgroup. It is straightforward to see that the element $d = [\frac{1+e_1}{\sqrt{2}}, \frac{1-e_1}{\sqrt{2}}]$ exchanges the two icosahedral orbits; $d: O(\Lambda_I) \leftrightarrow O(\Lambda_{II})$. One can see that the element $d$ represents a rotation around the first axis by $90^0$ and extends the group $A_4 \approx [T,\bar{T}]$ to the proper octahedral rotation group $S_4 \approx A_4 : C_2 \equiv \{[T,\bar{T}] \oplus T',\bar{T}']\}$. This proves that two mirror images of the icosahedron are transformed to each other by rotations therefore the icosahedron is not a chiral solid rather it is achiral. When two orbits of (28) and (29) are combined one obtains a quasi regular polyhedron which can be obtained as the orbit of the group $W(B_3)(1,\tau,0)$ [14]. The quasi regular polyhedron represented by the combined vertices of (28-29) is shown in Fig. 8. It consists of two types of faces, squares of side $\tau$ and isogonal hexagons of sides 1 and $\tau$.

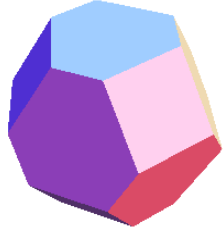

**Figure 8.** The quasi regular polyhedron represented by the vertices of (28-29).

Although we know that the dual of an icosahedron is a dodecahedron [6] here we show how the vertices of the dodecahedron can be obtained from the vertices of the icosahedron, say, from the vertices of $O(\Lambda_I)$ given in (28). We have to determine the centers of the planes in Fig.6. We can choose the vector $\omega_1$ as the vector representing the center of the face #1 because it is invariant under the rotation represented by $r_2 r_3$. In other words the triangle #1 is rotated to itself by a rotation around the vector $\omega_1$. With the same reason the center of the face #3 can be taken as the vector $\omega_3$. We note that the line joining these vectors is orthogonal to the vector $\Lambda_I$, namely, $(\omega_1 - \omega_3).\Lambda_I = 0$. The centers of the faces #2, #4 and #5 can be determined by averaging the vertices representing these faces:

$$b_2 = \frac{1}{3}(\Lambda_I + r_2 r_1 \Lambda_I + r_2 r_3 \Lambda_I), b_4 = \frac{1}{3}(\Lambda_I + r_1 r_2 \Lambda_I + r_1 r_3 \Lambda_I), b_5 = \frac{1}{3}(\Lambda_I + r_3 r_1 \Lambda_I + r_3 r_2 \Lambda_I) \quad . \quad (30)$$

Since we have the following relations among these three vectors $r_1 r_2 b_2 = b_4$, $r_3 r_2 b_4 = b_5$ and $r_3 r_2 b_2 = b_5$ they are in the same orbit under the group action $A_4(b_2) = A_4(b_4) = A_4(b_5)$. Therefore it is sufficient to work with one of these vectors, say, with $b_2$. In terms of the quaternionic units it reads $b_2 = \dfrac{\tau^2}{6}(\tau e_2 - \sigma e_3)$. A quick check shows that $b_2 - \omega_1$ is not orthogonal to the vector $\Lambda_I$, rather $(\lambda b_2 - \omega_1).\Lambda_I = 0$ provided $\lambda = 3\sigma^2$. Then we obtain three orbits

$$A_4(\lambda b_2) = \{\tfrac{1}{2}(\pm\tau e_1 \pm \sigma e_2),\ \tfrac{1}{2}(\pm\tau e_2 \pm \sigma e_3),\ \tfrac{1}{2}(\pm\tau e_3 \pm \sigma e_1)\} \qquad (31)$$

$$A_4(\omega_1) = \{\tfrac{1}{2}(e_1 + e_2 + e_3), \tfrac{1}{2}(-e_1 - e_2 + e_3), \tfrac{1}{2}(-e_1 + e_2 - e_3), \tfrac{1}{2}(e_1 - e_2 - e_3)\}$$

$$A_4(\omega_3) = \{\tfrac{1}{2}(-e_1 - e_2 - e_3), \tfrac{1}{2}(-e_1 + e_2 + e_3), \tfrac{1}{2}(e_1 - e_2 + e_3), \tfrac{1}{2}(e_1 + e_2 - e_3)\}.$$

Note that the last two orbits represent the vertices of two dual tetrahedra, when combined, represent a cube. These 20 vertices which decompose as three orbits under the tetrahedral group represent the vertices of a dodecahedron as shown in the Fig.9 which is also achiral solid. So far we have shown that, although, tetrahedron and icosahedron can be obtained as chiral solids there exists additional proper rotational group elements that convert them to their mirror images. Therefore they are not chiral solids.

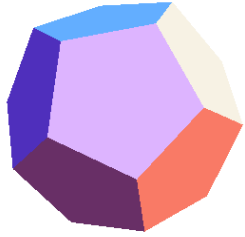

**Figure 9.** Dodecahedron represented by the vertices of (31).

Although our main topic is to study the chiral objects systematically using the Coxeter diagrams, here with a brief digression, we construct the pyritohedron, a non regular dodecahedron, made by 12 irregular pentagons. If we plot the solid represented by the orbit $A_4(\lambda b_2)$ in the first line of (31) we obtain an irregular icosahedron as shown in Fig.10.

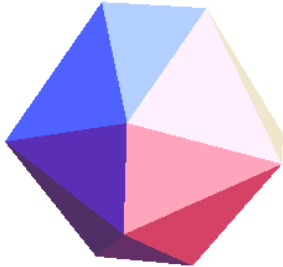

**Figure 10.** Irregular icosahedron represented by the vertices of the orbit $A_4(\lambda b_2)$.

Let us recall that the vector $\lambda b_2 = \frac{1}{2}(\tau\omega_1 - \omega_2 + \tau\omega_3)$ differs from the vector $\Lambda_I$ by the sign in front of $\omega_2$. In Fig. 11 we show the faces joining to the vector $\lambda b_2$.

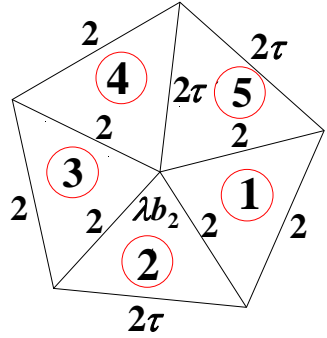

**Figure 11.** Faces of the irregular icosahedron joined to the vertex $\lambda b_2$.

There we see that two of the triangles are equilateral and the rest three are isosceles triangles. We determine the centers of the faces of this irregular icosahedron. The faces #1 and #3 can be represented again by the vectors $\omega_1$ and $\omega_3$ respectively. The centers of the faces of the #2, #4 and #5 can be determined, up to a scale factor, by averaging the vectors representing the vertices of the isosceles triangles. They can be obtained, up to a scale factor, as

$$d_2 \equiv (\tau - 2\sigma)e_2 - \sigma e_3, \quad d_4 \equiv (\tau - 2\sigma)e_3 + \sigma e_1, \quad d_5 \equiv (\tau - 2\sigma)e_3 - \sigma e_1. \tag{32}$$

These three vertices determine a plane which can be represented by its normal vector $D \equiv 5e_2 + (7+\sigma)e_3$ up to a scale factor. Now, we can determine the scale factor $\rho$ so that $(\rho d_2 - \omega_1).D = 0$ determines the five vertices lying in the same plane. We obtain $\rho = \frac{205 - 280\sigma}{1210}$. The particular edge represented by the vector $d_4 - d_5 = 2\sigma e_1$ leads to an orbit of size 6 given by $2\sigma\{\pm e_1, \pm e_2, \pm e_3\}$. This shows that Pyritohedral group transforms this type of edges to each other. The vertices of the pyritohedron are given by the set of quaternions:

$$\rho\{\pm(\tau - 2\sigma)e_1 \pm \sigma e_2, \ \pm(\tau - 2\sigma)e_2 \pm \sigma e_3, \ \pm(\tau - 2\sigma)e_3 \pm \sigma e_1\} \ ,$$

$$O(\omega_1) \oplus O(\omega_3) = \{\frac{1}{2}(\pm e_1 \pm e_2 \pm e_3)\} \tag{33}$$

which leads to the pyritohedron as shown in Fig.12. Its symmetry is represented by pyritohedral group $A_4 \times C_2 = \{[T,\bar{T}] \oplus [T,\bar{T}]^*\}$ of order 24. The 20 vertices of the pyritohedron lie in three orbits 20=12+4+4 as shown in (33).

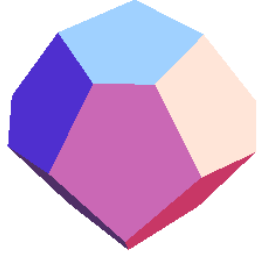

**Figure 12.** The pyritohedron consisting of irregular pentagonal faces.

A variety of pyritohedron can be constructed. If two orbits of the tetrahedron leading to the vertices of a cube determined are chosen to be the set of quaternions $(\pm e_1 \pm e_2 \pm e_3)$ then one can build the orbit of size-12 which depends on a single parameter. Indeed the following sets of 12 quaternions are invariant under the pyritohedral group $T_h$:

$$\{\pm ae_1 \pm be_2, \ \pm ae_2 \pm be_3, \ \pm ae_3 \pm be_1\} \tag{34}$$

where $a$ and $b$ are arbitrary real parameters. Here now, three vectors $ae_2 - be_3$, $ae_3 \pm be_1$ determine a plane whose normal can be represented by the vector $(a+b)e_2 + ae_3$. The condition that five points represented by the vectors $(e_1 + e_2 + e_3)$, $(-e_1 + e_2 + e_3)$ and $ae_2 - be_3$, $ae_3 \pm be_1$ are in the same plane determines that $b = a^2 - 2a$. Therefore the set of vertices of a pyritohedron has an arbitrary parameter and includes also dodecahedron and the rhombic dodecahedron [6], a Catalan solid, as members of the family for $a = \tau$ and $a = 2$ respectively. The pyritohedron is face-transitive since the normal vectors of the faces form an orbit of size 12 under the pyritohedral group. It is an achiral solid.

## 5  The snub cube derived from the orbit $O(\Lambda) = (W(B_3)/C_2)(a_1 a_2 a_3)$.

The snub cube is an Archimedean chiral solid. Its vertices and its dual solid can be determined employing the same method described in Sec.3 and Sec.4. The proper rotational subgroup of the Coxeter group $W(B_3) \approx S_4 \times C_2$ is the octahedral group $W(B_3)/C_2 \approx S_4$, isomorphic to the symmetric group of order 24. They can be generated by the generators $a = r_1 r_2$ and $b = r_2 r_3$ which satisfy the generation relation $a^3 = b^4 = (ab)^2 = 1$. Let $\Lambda = (a_1 a_2 a_3)$ be a general vector. The following sets of vertices form an equilateral triangle and a square respectively

$$(\Lambda, \ r_1 r_2 \Lambda, \ (r_1 r_2)^2 \Lambda), \qquad (\Lambda, \ r_2 r_3 \Lambda, \ (r_2 r_3)^2 \Lambda, \ (r_2 r_3)^3 \Lambda), \tag{35}$$

with the respective square of edge lengths $2(a_1^2 + a_1 a_2 + a_2^2)$ and $2(a_2^2 + \sqrt{2} a_2 a_3 + a_3^2)$. We have another vertex $r_1 r_3 \Lambda = r_3 r_1 \Lambda$ as shown in Fig.13.

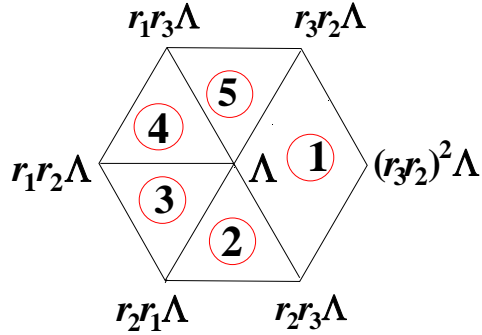

**Figure 13.** The vertices connected to the vertex $\Lambda$.

Similar to the arguments discussed in Sec.4 we obtain four equilateral triangles and one square sharing the vertex $\Lambda = (a_1 a_2 a_3)$ (see Fig. 13) provided the following equations are satisfied

$$(a_1^2 + a_1 a_2 + a_2^2) = (a_2^2 + \sqrt{2} a_2 a_3 + a_3^2) = (a_1^2 + a_3^2).$$

Factoring by $a_2^2$ and defining again $x = \dfrac{a_1}{a_2}$ and $y = \dfrac{a_3}{a_2}$ one obtains $y = \dfrac{x^2 - 1}{\sqrt{2}}$ and the cubic equation $x^3 - x^2 - x - 1 = 0$. This equation has one real solution $x \approx 1.8393$. Now the first orbit can be derived from the vector $\Lambda_I = a_2 (x \omega_1 + \omega_2 + y \omega_3)$ and its mirror image can be defined as $\Lambda_{II} = r_1 \Lambda_I = a_2 (x(\omega_1 - \alpha_1) + \omega_2 + y \omega_3)$. In terms of quaternionic units the vectors read

$$\Lambda_I = \frac{a_2 (x^2 + 1)}{2} (x e_1 + e_2 + x^{-1} e_3) \ , \quad \Lambda_{II} = \frac{a_2 (x^2 + 1)}{\sqrt{2}} (e_1 + x e_2 + x^{-1} e_3) \ . \qquad (36)$$

Deleting the overall scale factor in (36) the set of vectors constituting the orbits can be easily determined [6] as

$$O(\Lambda_I) = \{(\pm x e_1 \pm e_2 \pm x^{-1} e_3), \ (\pm x e_2 \pm e_3 \pm x^{-1} e_1), \ (\pm x e_3 \pm e_1 \pm x^{-1} e_2)\} \ ,$$

$$O(\Lambda_{II}) = \{(\pm e_1 \pm x e_2 \pm x^{-1} e_3), \ (\pm e_2 \pm x e_3 \pm x^{-1} e_1), \ (\pm e_3 \pm x e_1 \pm x^{-1} e_2)\} \ . \qquad (37)$$

The snub cubes represented by these sets of vertices are depicted in Fig.14. Note that no proper rotational symmetry exists which transforms these two mirror images to each other so that they are truly chiral solids.

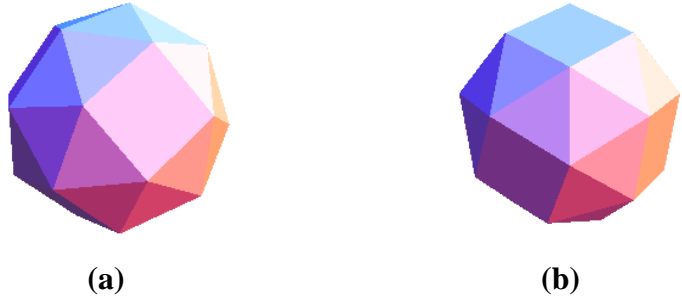

**(a)**          **(b)**

**Figure 14.** Two snub cubes (a) $O(\Lambda_I)$ and (b) $O(\Lambda_{II})$ (mirror image of each other).

One can combine the vertices of these two chiral solids in one solid which is achiral and it is depicted in Fig.15. This quasi regular solid can be obtained from the vector $\Lambda_I = a_2(x\omega_1 + \omega_2 + y\omega_3)$ by applying the octahedral group $W(B_3)(\Lambda_I)$.

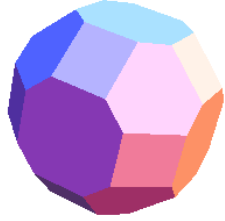

**Figure 15.** The quasi regular polyhedron consisting of two chiral orbits $O(\Lambda_I)$ and $O(\Lambda_{II})$.

The dual of the snub cube can be determined by determining the centers of the faces as shown in Fig. 13.

Similar arguments discussed in Sec. 4 can be used to determine the centers of the faces in Fig.13. The faces #1 and #3 are represented by the vectors $\omega_1$ and $\omega_3$ respectively. The vectors representing the centers of the faces #2, #4 and #5 can be determined and they lie in the same orbit under the proper octahedral group. The vector representing the center of the face #2 can be given, up to a scale factor, in terms of the quaternionic units as

$$c_2 = (2x+1)e_1 + e_2 + x^2 e_3 \ . \tag{38}$$

The scale factors multiplying the vectors $\lambda\omega_1, \omega_3$ and $\eta c_2$ can be determined as $\lambda = \dfrac{x}{\sqrt{2}}$ and $\eta = \dfrac{x^{-2}}{\sqrt{2}}$ when $\Lambda_I$ represents the normal of the plane containing these five points.

Then 38 vertices of the dual solid of the snub cube, the pentagonal icositetrahedron, are given in three orbits as follows

$$O(\lambda\omega_1) = \frac{x}{\sqrt{2}}\{\pm e_1, \pm e_2, \pm e_3\}$$

$$O(\omega_3) = \frac{1}{\sqrt{2}}(\pm e_1 \pm e_2 \pm e_3)$$

$$O(\eta c_2) = \frac{x^{-2}}{\sqrt{2}}\{[\pm(2x+1)e_1 \pm e_2 \pm x^2 e_3],[\pm(2x+1)e_2 \pm e_3 \pm x^2 e_1],[\pm(2x+1)e_3 \pm e_1 \pm x^2 e_2]\}$$

(39)

The pentagonal icositetrahedron is shown in Fig.16.

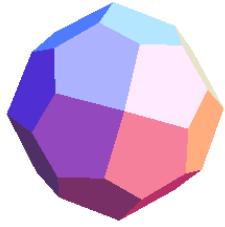

**Figure 16.** The pentagonal icositetrahedron, a Catalan solid, dual of the snub cube.

## 6  The snub dodecahedron derived from the orbit $O(\Lambda) = (W(H_3)/C_2)(a_1 a_2 a_3)$

The snub dodecahedron is the second Archimedean chiral solid. Its vertices and its dual solid can be determined employing the same method described in section 5. The proper rotational subgroup of the Coxeter group $W(H_3) \approx A_5 \times C_2$ is the icosahedral group $W(H_3)/C_2 \approx A_5$, which is the simple finite group of order 60 representing the even permutations of five letters. They can be generated by the generators $a = r_1 r_2$ and $b = r_2 r_3$ which satisfy the generation relation $a^5 = b^3 = (ab)^2 = 1$. Let $\Lambda = (a_1 a_2 a_3)$ be a general vector. The following sets of vertices form a pentagon and an equilateral triangle respectively

$$(\Lambda,\ r_1 r_2 \Lambda,\ (r_1 r_2)^2 \Lambda,\ (r_1 r_2)^3 \Lambda,\ (r_1 r_2)^4 \Lambda), \qquad (\Lambda,\ r_2 r_3 \Lambda,\ (r_2 r_3)^2 \Lambda), \qquad (40)$$

with the respective square of edge lengths $2(a_1^2 + \tau a_1 a_2 + a_2^2)$ and $2(a_2^2 + a_2 a_3 + a_3^2)$. We have another vertex $r_1 r_3 \Lambda = r_3 r_1 \Lambda$ which is depicted in Fig.17.

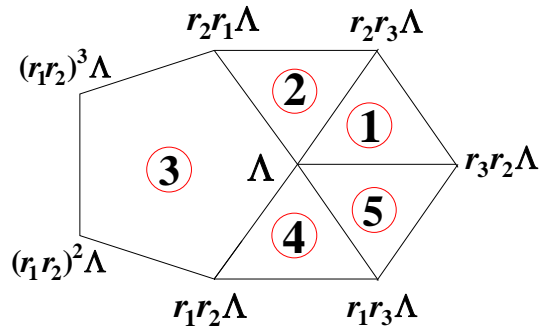

**Figure 17.** The vertices connected to the vertex $\Lambda$.

The only difference of this from the one in Fig.13 is that in the present case the face #1 is an equilateral triangle whose center is represented by the vector $\omega_1$ and the face #3 is a regular pentagon whose center is represented by the vector $\omega_3$. Assuming that the face #1 face #2, face #4 and face #5 are equilateral triangles which lie in the same orbit of size 60 one obtains the following equations:

$$(a_1^2 + \tau a_1 a_2 + a_2^2) = (a_2^2 + a_2 a_3 + a_3^2) = (a_1^2 + a_3^2). \quad (41)$$

Factoring by $a_2^2$ and defining $y = \dfrac{a_1}{a_2}$ and $x = \dfrac{a_3}{a_2}$ one obtains $y = \dfrac{x^2-1}{\tau} = \dfrac{\tau}{x-1}$ and the cubic equation is $x^3 - x^2 - x - \tau = 0$. This equation has the real solution $x \approx 1.94315$. Now the first orbit can be derived from the vector $\Lambda_I \equiv a_2(\dfrac{\tau}{x-1}\omega_1 + \omega_2 + x\omega_3)$ and its mirror image can be defined as $\Lambda_{II} \equiv r_1 \Lambda_I = a_2(\dfrac{\tau}{x-1}(\omega_1-\alpha_1) + \omega_2 + x\omega_3)$. In terms of the quaternionic units these vectors read

$$\Lambda_I = \dfrac{a_2}{\sqrt{2}}[\sigma(x^2-1)e_1 - xe_2 + (1-\tau x^3)e_3]\, ,\, \Lambda_{II} = \dfrac{a_2}{\sqrt{2}}[-\sigma(x^2-1)e_1 - xe_2 + (1-\tau x^3)e_3]\, .\, (42)$$

Two snub dodecahedra obtained using these vectors are shown in Fig.18 (a) and (b). One can combine the vertices of these two chiral solids in one solid which is achiral and it is depicted in Fig.18 (c). This quasi regular solid (quasi regular great rhombicosidodecahedron) can be obtained from the vector $\Lambda_I = a_2(\dfrac{\tau}{x-1}\omega_1 + \omega_2 + x\omega_3)$ by applying the icosahedral group $W(H_3)(\Lambda_I)$.

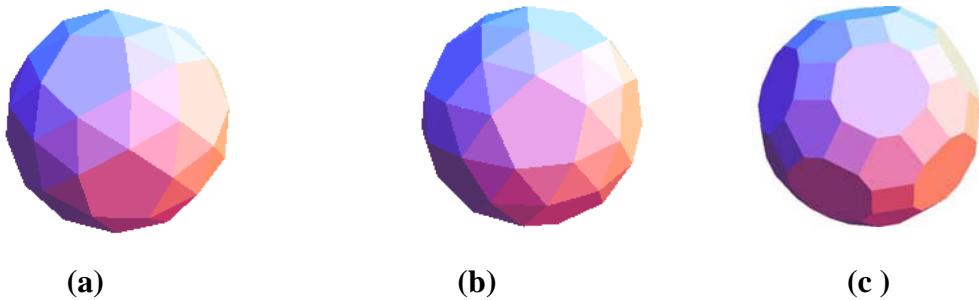

(a)  (b)  (c)
**Figure 18.** The snub dodecahedron (a) $O(\Lambda_I)$ and (b) its mirror image $O(\Lambda_{II})$, (c) quasi regular great rhombicosidodecahedron.

The vertices of the dual solid (pentagonal hexecontahedron) of the snub dodecahedron represented by $O(\Lambda_I)$ can be given as the union of three orbits of the group $W(H_3)/C_2 \approx A_5$. The orbit $O(\omega_1)$ consists of 20 vertices of a dodecahedron. The second orbit consists of 12 vertices of an icosahedron $\lambda O(\omega_3)$ where

$$\lambda = \frac{-3\sigma x^2 + \tau x + 2 + \sigma}{x^2 + (2+\sigma)x + 1} . \tag{43}$$

The third orbit $\rho O(c_2)$ involves the vertices including the centers of the faces #2, #4 and #5 where the vector $\rho c_2$ is given by

$$\rho c_2 = \frac{1}{\sqrt{2}} \frac{3\tau x^2 + \tau^3 x + \tau + 2}{(21\tau+20)x^2 + (21\tau+17)x + 21\tau+11} \{[(2\sigma-1)x^2 - \tau x - 1]e_1 \\ + (-x^2 + 3\tau x + 3)e_2 - \tau^3(x^3 + \sigma)e_3\} \tag{44}$$

Applying the group $A_5 = [I, \bar{I}]$ on the vector $\rho c_2$ one generates an orbit of size 60. The 92 vertices consisting of these three orbits constitute the dual solid (pentagonal hexecontahedron) of the snub dodecahedron represented by the orbit $O(\Lambda_1)$.
The pentagonal hexecontahedron is shown in Fig. 19.

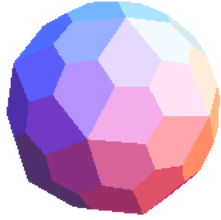

**Figure 19.** The pentagonal hexacontahedron, the dual of the snub dodecahedron.

The pentagonal hexecontahedron is one of the face transitive Catalan solid which has 92 vertices, 180 edges and 60 faces.

### 7 Concluding Remarks

In this work we presented a systematic construction of the chiral polyhedra, the snub cube, snub dodecahedron and their duals using proper rotational subgroups of the octahedral group and the icosahedral group. We used the Coxeter diagrams $B_3$ and $H_3$ respectively. Employing the same technique for the diagrams $A_1 \oplus A_1 \oplus A_1$ and $A_3$ we obtained the vertices of tetrahedron and icosahedrons which are not the chiral solids because they can be transformed to their mirror images by the proper rotational subgroup of the octahedral group.
As a by-product we also constructed the orbit of the pyritohedron using the pyritohedral group which is the symmetry of the iron pyrits.
This method can be extended to the higher dimensional Coxeter groups to determine the chiral polytopes. For example, the snub 24-cell, a chiral polytope in the 4D Euclidean space can be determined using the $D_4$ Coxeter diagram [15].